# New Misfit-Layered Cobalt Oxide (CaOH)$_{1.14}$CoO$_2$

Mitsuyuki Shizuya [1], Masaaki Isobe [1,*], Yuji Baba [2], Takuro Nagai [3], Minoru Osada [2], Kosuke Kosuda [4], Satoshi Takenouchi [4], Yoshio Matsui, [2,3] and Eiji Takayama-Muromachi [1,2]

[1] *Superconducting Materials Center, National Institute for Materials Science (SMC/NIMS), 1-1 Namiki, Tsukuba, Ibaraki 305-0044, JAPAN*

[2] *Advanced Materials Laboratory, National Institute for Materials Science (AML/NIMS), 1-1 Namiki, Tsukuba, Ibaraki 305-0044, JAPAN*

[3] *High Voltage Electron Microscopy Station, National Institute for Materials Science (HVEMS/NIMS), 1-1 Namiki, Tsukuba, Ibaraki 305-0044, JAPAN*

[4] *Materials Analysis Station, National Institute for Materials Science (MAS/NIMS), 1-1 Namiki, Tsukuba, Ibaraki 305-0044, JAPAN*

We found a new cobalt oxide (CaOH)$_{1.14}$CoO$_2$ by utilizing the high-pressure technique. X-ray and electron diffraction studies revealed that the compound has layer structure which consists of CdI$_2$-type CoO$_2$ layers and rock-salt-type double CaOH atomic layers. The two subcells have incommensurate periodicity along the *a*-axis, resulting in modulated crystal structure due to the inter-subcell interaction. The structural modulation affects carrier conduction through the potential randomness. We found that the two-dimensional (2-D) variable-range hopping (VRH) regime with hole conduction is dominant at low temperature for this compound, and that the conduction mechanism undergoes crossover from the 2-D VRH regime to thermal activation-energy type one with increasing temperature. Based on the experimental results of resistivity, thermoelectric power, magnetic susceptibility and specific heat measurements, we suggested a possible electronic-band structure model to explain these results. The cobalt $t_{2g}$-derivative band crosses Fermi energy level near the band edge, yielding small finite density of localized states at the Fermi level in the band. The observed resistivity, Seebeck coefficient, large Pauli paramagnetic component in the magnetic susceptibility and comparatively small Sommerfeld constant in the specific heat are principally attributed to the holes in the $t_{2g}$-derivative band. We estimated the Wilson ratio to be about 2.8, suggesting the strong electron correlation realized in this compound.

# I. INTRODUCTION

Layered cobalt oxides have recently attracted much attention from solid-state physicists and chemists because of various unusual physical properties that they possess. In particular, unconventional superconductivity found in $Na_{0.35}CoO_2 \cdot 1.3H_2O$ [1] much stimulates our intellectual curiosity, because origin of the superconductivity seems to be associated with rather unusual spin-triplet-type electron pairing with ferromagnetic spin-fluctuation. [2] Another most interesting characteristic feature is large thermoelectric power realized in the layered cobalt oxides such as $\gamma$-$Na_{0.5}CoO_2$ [3], $[Ca_2CoO_3]_{0.62}CoO_2$ [4], and $[Bi_{0.87}SrO_2]_2[CoO_2]_{1.82}$. [5] These compounds have comparatively large Seebeck coefficients $S$ in spite of their metallic resistivity $\rho$, resulting in large power factors $S^2/\rho$. In general, a large thermoelectric figure of merit $Z$ ($=S^2/\rho\kappa$, $\kappa$: thermal conductivity) is an important factor from the point of view of application of compounds to high-performance thermoelectric device. Moreover, many other interesting features, *e.g.* large magneto-resistance [4,6,7], magnetic anomaly [8,9], and spin-density wave [10], have been observed in the layered cobalt oxides. These unusual physical properties seem to be caused by strong Coulomb interaction and collective excitation of electrons in a cobalt-oxygen hybridized narrow band. It is significant for progress in solid-state physics to understand the origin of the quantum phenomena due to the strong electron correlation.

In order to understand the nature of the electron correlation in the layered cobalt oxides, it is important to find a new class of materials with more fascinating physical properties such as the unconventional superconductivity. The hope is that the search for new materials will lead to an improved understanding of the unusual physical properties and ultimately to discovery of novel functions in the materials. The layered cobalt oxides have common features in the crystal structure. The crystal structure consists of $CdI_2$-type conducting $CoO_2$ layers and insulating blocking layers. Accordingly, if we replace the blocking layer with any other types of structure with keeping the $CoO_2$ layer on, we can obtain a new class of materials. This idea is similar to that for material design of the homologous series of high-$T_c$ cuprate superconductors. [11,12] Thus far, a lot of high-$T_c$ cuprate superconductors with various structure types of blocking layers have been found on the basis of the similar idea for the material design, and the discovery of the new materials have much supported further understanding of physics and chemistry in the high-$T_c$ superconductivity. Our original idea in this study is to apply the same guiding principle for the material design to new phase search for the layered cobalt oxides.

The layered cobalt oxides can be classified by the number of atomic layers $n$ in the blocking layer. Figure 1 shows a schematic representation of crystal structures of a series of the layered cobalt oxides. For instance, $Na_xCoO_2$ ($x$=0.3~0.7) has a single atomic layer of sodium between the $CoO_2$ layers, and this can be regarded as the $n$=1 member of the series. For the other compounds, $[Ca_2CoO_3]_{0.62}CoO_2$ [4, 13] and $[Bi_{0.87}SrO_2]_2[CoO_2]_{1.82}$ [14] have triple and quadruple atomic layers in the blocking layers with the rock-salt type structure; namely, these are the $n=3$ and $n=4$ members of the series, respectively. However, only the double atomic-layer compound, the $n=2$ member of the series, has not yet been reported at least as far as we know. This is "missing link" in the series of the layered cobalt oxides. The aim of the present work was to prepare the $n=2$ member of the series of the layered cobalt oxides by utilizing high-pressure synthesis technique, and to complete the series. Generally speaking, high-pressure condition is favorable for stabilizing the phases which include closest-packing structure such as the $CoO_2$ layer block. Indeed, we carried out phase search experiments under high pressure, and recently discovered a new compound having double atomic layers in the blocking layer.

In this paper, we report on the novel cobalt calcium hydroxide $(CaOH)_{1.14}CoO_2$. X-ray and electron diffraction studies revealed that this compound is a composite crystal which consists of interpenetrated two subsystems of the $CdI_2$-type $CoO_2$ layers and the rock-salt-type double CaOH atomic layers. This structure is similar to that in a kind of layered sulfides, $(MS)_xTS_2$ ($M$=Sn, Pb, Bi or lanthanides; $T$=Nb, Ta, Ti, V or Cr). [15-18] We found that the present compound is electrically insulating due to insufficiency of doped carriers; however, the resistivity, thermoelectric power, magnetic susceptibility and specific-heat studies indicated that the compound has semimetal-like electronic-band structure with localized states. It implies that this compound is a possible candidate for the prototype material of a new superconductor following $Na_{0.35}CoO_2 \cdot 1.3H_2O$.

## II. EXPERIMENT

A polycrystalline sample was prepared by means of solid-state reaction using high-pressure synthesis technique. The sample was made from a mixture of starting materials, $Co_3O_4$ (99.9%), $CaO_2$ (99%), CaO and $Ca(OH)_2$. [19] The reagents with a molar ratio of Ca: Co: O: H = 1.117: 1: 3.14: 1.049 were mixed using an agate mortar in a glove box filled with dry Ar gas. [20] The mixture was sealed into a gold capsule, and then allowed to react in a flat-belt-type high-pressure apparatus under 6 GPa at 1373–1473 K for 1 hour, followed by quenching to room temperature before releasing the pressure.

Purity of the product was carefully checked by powder x-ray diffraction. We confirmed that the as-sintered product contains a small amount of $Ca(OH)_2$ as an impurity phase. To remove the $Ca(OH)_2$, some fractions of the product were roughly pulverized and washed with ion-exchanged water (200 ml per 200–300 mg sample powder) for 5 minutes using an ultrasonic cleaner. The wash process was repeated three times with exchanging the water for clean one. After the wash, the powder was dried at 423 K in air. Hereafter, we abbreviate the as-sintered bulk ceramics to "bulk sample" or "as-grown sample" and the washed powder to "washed sample".

The cation ratio of the high-pressure phase in the sample was determined by electron probe microanalysis (EPMA) using a wavelength-dispersive x-ray spectrometer (JEOL JXA-8500F) with an acceleration voltage of 15 kV. In EPMA, a small ceramic specimen of the bulk sample was well polished using a 0.3 μm alumina lapping film to obtain a flat surface, and several relative large grains were selected and analyzed. The average grain size of the phase was about 100 μm × 30 μm.

Hydrogen content in the washed sample was determined by infrared (IR) absorption spectroscopic analysis with a carbon/hydrogen analyzer (LECO RC-412). In the measurement, as temperature increases, hydrogen atoms in the sample react with oxygen atoms in the phase and/or carrier oxygen gas, and simultaneously yield $H_2O$ vapor. Intensity of the IR-absorption spectrum due to the vaporized $H_2O$ was recorded during the heating. The amount of $H_2O$ in the sample was determined by calibrating the intensity data using a standard material $Na_2WO_4 \cdot 2H_2O$. To avoid detecting the water adhering onto grain surfaces and/or boundaries, the samples were pre-heated at 423 K before the measurement.

The content of cobalt atoms in the washed sample was determined by inductively coupled plasma atomic emission spectrometry (ICP-AES), after dissolving the sample powder in hydrochloric acid. The oxidation state of the cobalt ion was determined by redox titration. The washed sample was dissolved in sulfuric acid containing an excess of sodium oxalate, $(COONa)_2$, as a reducing agent. The residual $(COONa)_2$ was titrated against an aqueous solution of potassium permanganate ($KMnO_4$) to reduce the oxidation state of the cobalt ions.

Powder x-ray diffraction (XRD) data were collected at room temperature using a diffractometer (Rigaku, RINT2200HF-ULTIMA) equipped with Bragg-Brentano geometry and CuKα radiation. Lattice constants were determined by the least-squares method. Crystal structure was analyzed by computer simulation of the XRD pattern using the software PREMOS91 on the basis of superspace-group symmetry approach.[21] Transmission electron microscopy observations were carried out using a microscope (Hitachi H-1500) operating at 820 kV. The bulk sample was pulverized in an agate mortar, and the powder was dispersed in $CCl_4$ using an ultrasonic cleaner. The supernatant fluid containing fine powder was collected and dropped onto carbon micro-grids for the observation.

Raman experiments were carried out in a backward micro-configuration to study the bonding state of hydrogen in the phase. An $Ar^+$ laser (1 mW, 514.5 nm line) beam with a 2 μm-diameter spot was focused on individual grains in the washed sample at room temperature. The scattered light was recorded using a subtractive triple liquid-$N_2$-cooled spectrometer (T64000, HORIBA Jobin-Yvon) equipped with a charge-coupled-device (CCD) detector.

Magnetic data were collected for pulverized sample using a superconducting quantum interference device (SQUID) magnetometer (Quantum Design, MPMS) on cooling at the magnetic field of 1000 Oe between 2 K and 300 K. Specific heat was measured by a relaxation method using a commercial apparatus (Quantum Design, PPMS) between 2 K and 300 K. The weight of the bulk sample used for the measurement was 16.97 mg.

Seebeck coefficient was measured by a thermal transport option (TTO) using the PPMS with a four-probe configuration. The sample size was 1.4×2.0×5.3 mm$^3$, and the distance between the two temperature/voltage terminals was 2.3 mm. Copper wires were attached onto the polished sample surface via evaporated thin gold film using silver paste in order to make fine ohmic contact. Data were collected by a continuous mode on cooling of 0.3 K/min. Temperatures at the two temperature/voltage terminals were detected by thermometers (Cernox 1050), and the temperature difference between the two terminals was controlled within 3 % of the measurement temperature. Electrical resistivity was measured by the conventional four-probe AC method simultaneously with the thermoelectric-power measurement. AC current of 0.5–0.01 mA with 60–300 Hz was applied to the sample.

## III. RESULTS AND DISCUSSION

### A. Characterization

Figure 2 shows x-ray diffraction (XRD) patterns of the washed-sample; the inset shows a whole $2\theta$ range pattern, while the main panel indicates enlargement of the pattern. We confirmed that the washed sample does not contain any trace of the Ca(OH)$_2$ impurity phase as shown by arrows in the main panel of Fig. 2. It indicates that the Ca(OH)$_2$ phase was entirely removed from the sample by the water-wash process. Any other difference was not observed in the XRD patterns before and after the wash process, suggesting that the main phase does not react with water.

Three intense diffraction peaks with $d$-values of ≈8.7×$n$ Å ($n$=1, 2, 3) were observed in the low $2\theta$ angle range in Fig. 2. Intensity of the diffraction peaks with $2\theta \geq 35°$ is much weaker than those of the lower $2\theta$ angle peaks; the intensity is less than one fortieth of the strongest peak intensity. This is due to strong preferred orientation. It suggests that the present phase has layer structure with periodicity of $d$≈8.7 Å along the direction perpendicular to the layer, and that a crystal grain easily cleaves into many plate-like crystals by the x-ray sample preparation.

In Fig. 2, all diffraction peaks can be systematically indexed by sets of four integers, $hklm$, by assuming that the present phase is a composite crystal having a (3+1)-dimensional structure with two $c$-centered orthorhombic subsystems with common $b$ (=4.9228(4) Å) and $c$ (=17.275(1) Å), and different $a$: $a_1$=2.8238(2) Å for the subsystem-1 and $a_2$=4.944(2) Å for the subsystem-2. The $a_1$, $a_2$ and $b$ parameters are near to the lattice constants of other misfit-layered cobalt oxides: *e.g.*

$a_1$=2.8238(2) Å, $a_2$=4.5582(2) Å, $b$=4.8339(3) Å for [Ca$_2$CoO$_3$]$_{0.62}$CoO$_2$ [13] and $a_1$=2.8081(5) Å, $a_2$=5.112(1) Å, $b$=4.904(1) Å for [Bi$_{0.87}$SrO$_2$]$_2$[CoO$_2$]$_{1.82}$. [14] It suggests that the intra-layer structure in the present compound is similar to those of the reference compounds. For the subsystem-1, the $b$ is nearly equal to $\sqrt{3}\,a_1$, and unit cell has a $c$-centered Bravais lattice. The lattice points form an approximately equilateral triangle lattice with hexagonal symmetry in the $a$-$b$ plane. This indicates that the subsystem-1 consists of CoO$_2$ layers with CdI$_2$-type structure. For the subsystem-2, the $a_2$ is nearly equal to $b$, suggesting that the unit lattice has approximately tetragonal symmetry. These parameters, $a_2$ and $b$, are close to twice of the Ca–O distance 2.4 Å estimated by Shannon's ionic radii: 1.0 Å for Ca$^{2+}$ (6-fold coordination) and 1.4 Å for O$^{2-}$ (6-fold coordination). [22] This indicates that the subsystem-2 consists of CaO blocks with rock-salt-type structure.

The $c/2$ is 8.638 Å, which corresponds to the nearest-neighbor distance between the CoO$_2$ layers. This length is comparable with the $c$ parameters of [Ca$_2$CoO$_3$]$_{0.62}$CoO$_2$ ($c$=10.8436(7) Å). [13] The difference in these values is due to the different number of stacking layers in the rock-salt-type structure block. [Ca$_2$CoO$_3$]$_{0.62}$CoO$_2$ has triple layers, [CaO]-[CoO]-[CaO], intervening between two CoO$_2$ layers along the $c$-axis. Difference of the [CoO$_2$] inter-layer distances between the present compound and [Ca$_2$CoO$_3$]$_{0.62}$CoO$_2$ is 2.206 Å, which is relatively close to the [CoO] mono-layer thickness 1.945 Å estimated using Shannon's ionic radii: 0.545 Å for Co$^{3+}$ (6-fold coordination, low-spin state) and 1.4 Å for O$^{2-}$ (6-fold coordination). [22] It suggests that the present compound has double [CaO] layers for the rock-salt-type structure block.

The $a_1/a_2$ (=0.5711(4)) is the ratio of incommensurate periodicity along the $a$-axis between the [CoO$_2$] layer and the [CaO] block. The twice of this value corresponds to the Ca/Co ratio in composition of the present compound. Actual Ca/Co cation ratio determined by EPMA was 1.14, which is consistent with the value estimated from the structural study.

By the x-ray diffraction study, the composition of the present compound is expected to be [Ca$_2$O$_2$]$_{0.57}$[CoO$_2$], namely, (CaO)$_{1.14}$CoO$_2$, if we leave hydrogen atoms out of consideration. However, this is not exact, and we have to take account of presence of hydrogen atoms for the actual composition. We measured the actual amount of hydrogen in the present phase by an IR-absorption spectroscopic analysis. The resultant total amount of H$_2$O vaporized was 6.9 wt%, which corresponds to 1.2 hydrogen atoms per (CaO)$_{1.14}$CoO$_2$. This value is very close to the amount of oxygen atoms 1.14 in the CaO block, suggesting that the almost all hydrogen atoms are situated in the CaO block. We conceive that the hydrogen atoms bond with the oxygen atoms in the CaO block, yielding OH$^-$ ions. The oxidation state of the cobalt ion determined by redox titration was +2.9. This value is consistent with the ideal cobalt valence +2.86 estimated from the composition (CaOH)$_{1.14}$CoO$_2$, where we assumed the charge neutrality and the cation valences: Ca$^{2+}$, OH$^-$ and O$^{2-}$. We therefore concluded that the phase composition (CaOH)$_{1.14}$CoO$_2$ is almost correct.

We observed the misfit-layered structure by transmission electron microscopy. Figures 3

(a), (b) and (c) show electron diffraction (ED) patterns projected along the [001], [010] and [100] directions, respectively. Fig. 3 (a) is a typical ED pattern on $a^*$–$b^*$ section, which was frequently observed because the fraction of the crystal easily cleaves along the $a$–$b$ plane. This pattern is essentially identical to those observed in other misfit-layered cobalt oxides such as [$Ca_2CoO_3$]$_{0.62}CoO_2$ [13] and [$Bi_{0.87}SrO_2$]$_2$[$CoO_2$]$_{1.82}$.[14] In Figs. 3, fundamental reciprocal-lattice vectors of the two subsystems are shown by arrows. The all reflections, both the main and satellite reflections, can be assigned by a linear combination of four unit vectors, $a_1^*$, $b^*$, $c^*$, $a_2^*$, and four integers, $hklm$, with a reciprocal-lattice vector, $q$ ($d=1/|q|$), given by $q = ha_1^* + kb^* + lc^* + ma_2^*$. The $a_2^*/a_1^*$ obtained is about 0.57, which is consistent with the $a_1/a_2$ (=0.5711(4)) value obtained by the x-ray diffraction study. Reflection conditions for the fundamental spots are $h+k=2n$ for $hkl0$ and $k+m=2n$ for $0klm$, which indicates that each subsystem has a $c$-centered Bravais lattice. The other reflection conditions are as follows: $h+k=2n$ for $hk00$, $k+m=2n$ for $0k0m$, $h, l=2n$ for $h0l0$, $l, m=2n$ for $00lm$, and $k=2n$ for $0kl0$. Resultant possible space groups of the fundamental structures for the two subsystems are, therefore, $Cmcm$ (No. 63), $Cmc2_1$ (No. 36) and $C2cm$ (No. 40).[23]

Figure 4 shows a high-resolution transmission electron microscopy (HRTEM) image projected along the direction perpendicular to the $c$-axis. This clearly indicates the layer stacking along the $c$-axis in structure with the double CaOH atomic layers sandwiched by the two $CoO_2$ layers. The distance between the nearest-neighbor $CoO_2$ layers is about 8.7 Å, which is in agreement with the $c/2$ value (=8.638 Å) obtained by the x-ray diffraction study.

Based on the symmetry study in the electron diffraction, we here propose a possible crystal structure model of the fundamental crystal structure as shown in Fig. 5. This structure model was designed on the basis of the assumption that both the subsystems have symmetry of the space group $Cmc2_1$ (No. 36). Note that the space groups $Cmcm$ (No. 63) and $C2cm$ (No. 40) can not generate rock-salt-type structure for the subsystem-2. The possible combination of space groups is therefore limited to $Cmcm$(63)–$Cmc2_1$(36), $Cmc2_1$(36)–$Cmc2_1$(36) and $C2cm$(40)–$Cmc2_1$(36) for the subsystem-1 and subsystem-2. In the composite crystal, each substructure is modulated owing to the interaction between subsystems. Four-dimensional structure analysis technique is necessary for determining the conclusive superspace group and details of the modulated structure realized in this compound. In the present study, in advance of the precise analysis of the modulated crystal structure, we simulated the x-ray diffraction pattern using a computer program PREMOS91. Figure 6 shows an x-ray diffraction pattern simulated on the basis of the crystal structure model proposed in Fig. 5. This clearly indicates that the calculated diffraction pattern is considerably similar to the observed one, and that the proposed crystal structure model is close to the actual crystal structure.

From the chemical composition, it is most likely that hydrogen atoms exist in the rock-salt-type structure block as OH⁻ ions. We studied the bonding state between the oxygen and hydrogen atoms by Raman scattering experiments. Figure 7 shows a Raman spectrum for the washed sample. The peaks with the wave number below 800 cm$^{-1}$ are attributed to lattice vibration of ions, while the hump between 800 cm$^{-1}$ and 1500 cm$^{-1}$ are due to the second harmonic-generation of the lattice vibration. The peaks with the wave number above 2500 cm$^{-1}$ are related to molecular vibration of the –OH groups. A broad intense peak was observed in the wave-number range between 3000 cm$^{-1}$ and 3700 cm$^{-1}$. This peak is due to O–H stretching vibration mode in the −OH groups. No peak was observed around 1600 cm$^{-1}$. It indicates that the OH$^-$ ions are dominant for the hydrogen-oxygen bonding state in the present compounds, because the other possible bonding states, *e.g.* H$_2$O molecule and H$_3$O$^+$, should cause Raman scattering peaks around 1600 cm$^{-1}$ if they are included in this phase.[24]

The broad peak has the maximum at 3230–3380 cm$^{-1}$. These frequencies are very low and the peak width is very broad as compared with the typical peak frequency $\nu_{O-H} \approx 3600$ cm$^{-1}$ and peak width $\Delta \approx 10$ cm$^{-1}$ for general hydroxides such as Co(OH)$_2$.[25] The frequency shift indicates that the O–H binding energy in the OH$^-$ ions is weakened by interaction between the hydrogen and other ions. This interaction probably arises from hydrogen-bonding between the hydrogen and oxygen ions in the CoO$_2$ layer. It implies that the hydrogen atoms in the OH$^-$ ions are directed toward the nearest-neighbor oxygen ion in the CoO$_2$ layers. The peak broadening seems to be related to the structural incommensurability between the two subsystems. The inter-atomic distance between the OH$^-$ ion and the nearest-neighbor oxygen ion in the CoO$_2$ layer, *i.e.* hydrogen bonding distance, markedly varies depending on the each site position in structure. The strength of the OH$^-$ binding energy is affected by the structural modulation through the change of the hydrogen bonding distance. As a result, various strength of the OH$^-$ binding energy exists in the whole of crystal. This effect may cause the peak broadening of the O–H molecular vibration.

A small sharp peak which overlaps with the broad peak was observed at 3077 cm$^{-1}$. This peak position is very close to the Raman scattering frequency due to H$_3$O$^+$ ions, which were observed in the BLH phase of Na$_x$CoO$_2$•$y$H$_2$O.[26] However, this peak is negligibly small as compared with the integral intensity of the broad peak assigned to the OH$^-$ ion. We conceive that this sharp peak is extrinsic to the present phase.

**B. Physical properties**

Figure 8 (a) shows temperature dependence of electrical resistivity of (CaOH)$_{1.14}$CoO$_2$. A thermally activated behavior was observed over the temperature range measured. The resistivity at room-temperature is approximately $3 \times 10^1$ Ωcm, which is still two or three order larger in magnitude than the metallic limit for polycrystalline samples. The resistivity change looks to be semiconductor-like behavior; however, it does not keep to a simple activation-energy type mechanism through the temperature range. We found that the resistivity below about 270 K obeys the variable-range hopping (VRH) regime with the formula as follows:[27]

$$\rho(T) = \rho_0 \exp\left(\frac{T_0}{T}\right)^\nu, \qquad (1)$$

where the exponent $\nu$ is a fractional value between 1/4 and 1, and it depends on the dimensionality, electronic-band structure near the Fermi energy level ($E_F$), and Coulomb correlations between electrons.[27-29]

To determine the accurate $\nu$ values for the present compound, we analyzed the resistivity data using the following formula derived from Eq. (1):

$$\ln\left(-\frac{\partial \ln \rho}{\partial \ln T}\right) = (\ln \nu + \nu \ln T_0) - \nu \ln T \qquad (\nu \neq 0). \qquad (2)$$

In Fig. 8 (b), we plotted the $\ln[-\partial(\ln \rho)/\partial(\ln T)]$ vs. $\ln T$, where the slope of the fit gives $-\nu$. Three temperature ranges with different $\nu$ are clearly defined: (i) $\nu$=0.05 for 80–120 K, (ii) $\nu$=0.30 for 130–270 K, and (iii) $\nu$=0.82 for 300–360 K.[30] In the temperature range between 130 K and 270 K, the resistivity is well described by Eq. (1) with $\nu$=0.30. This $\nu$ value is approximately equal to 1/3 that is expected in case of the two-dimensional (2-D) VRH conduction.[27] We therefore concluded that the 2-D VRH regime is essentially dominant in this temperature range. The 2-D VRH regime observed in the temperature range probably originates in Anderson localization which is caused by potential randomness due to local inhomogeneity of chemical composition and/or structural modulation in the $CoO_2$ layers. This suggests that the compound has finite density of states $N(E)$ at the Fermi energy level $E_F$ and the carriers around $E_F$ are localized owing to the potential randomness. It implies that the electronic structure in the ground state is qualitatively near to what is expected in metals and not in semiconductors.

As temperature increases, the slope $\nu$ varies from 0.30 to 0.82 via narrow temperature range between 270 K and 300 K. It suggests occurrence of a thermally-induced crossover of the conduction mechanism. Above 300 K, the $\nu$ value approaches to 1, suggesting that the conductivity is associated with the activation-energy type conduction mechanism. The activation energy $\Delta E$ was estimated to be about 0.1 eV by fitting the resistivity data in the temperature range with the conventional formula: $\rho(T) = \rho_0 \exp(\Delta E / k_B T)$. We conceive that the $\Delta E$ value corresponds to the energy for thermal excitation of carriers from the $E_F$ to a mobility edge in the conduction or valence band.

Below 130 K, the resistivity gradually strays from the 2-D VRH regime with decrease in temperature. In the temperature range between 80 K and 120 K, the slope $\nu$ is kept near zero, where the resistivity approximately follows the formula: $\rho(T) = \rho_0 T^{-\alpha}$. Similar behavior is sometime observed in polycrystalline samples of other transition-metal oxides; however, the origin of such the conduction mechanism is not clear at least in the present compound. Perhaps, this may be related to generation of tunnel current between the $CoO_2$ layers and/or the effect of grain-boundary scattering in the polycrystalline sample.

The 2-D VRH regime was observed also in thermoelectric power for this compound. Figure 9 shows temperature dependence of the Seebeck coefficient $S$ of $(CaOH)_{1.14}CoO_2$. The $S$ has a positive sign through the whole temperature range measured, indicating that the majority carrier type is hole. The absolute value of $S$ at room-temperature is about 330 μV/K, which is as large as those of typical band-gap semiconductors such as Si. [31] However, the observed temperature dependence is not typical semiconductor-like ($S \propto 1/T$) through the temperature range, but rather metal-like ($S \propto T$) at low temperature. It has been theoretically and experimentally confirmed that the metallic behavior of $S$ occurs in case of VRH regime. [32-35] According to Mott $et\ al.$ [27], the Seebeck coefficient $S$ in the VRH regime is written by

$$S(T) = \frac{1}{2}\frac{k_B}{e}\frac{W^2}{k_B T}\left(\frac{\partial \ln N(E)}{\partial E}\right)_{E=E_F}, \qquad (3)$$

where $k_B$ is Boltzmann constant, $e$ is unit charge of electrons, and $W$ is hopping energy. Brenig $et\ al.$ defined the hopping energy in case of 2-D VRH regime as follows [35]:

$$W = \frac{1}{3}k_B T \left(\frac{T_0}{T}\right)^{\frac{1}{3}}. \qquad (4)$$

Substituting Eq. (4) into Eq. (3), we can obtain the form, $S \propto T^{1/3}$, in case of 2-D VRH regime. In the present compound, this 1/3rd-power law was clearly observed in the temperature range between 130 K and 220 K as shown in the inset of Fig. 9. The temperature range determined by the thermoelectric-power measurement (130–220 K) is nearly equal to that by the resistivity measurement (130–270 K).

Above 220 K, temperature dependence of the thermoelectric power strays out of the 2-D VRH conduction. The $S$ reaches the maximum at about 250 K, and then it decreases with increasing temperature above 250 K. It suggests that at 250 K the dominant electronic transport mechanism changes from the 2-D VRH regime to thermal-activation type band conduction by carrier excitation with excitation energy $\Delta E$. In this case, the Seebeck coefficient $S$ can be expressed by

$$S = \frac{k_B}{e}\left(\frac{\Delta E}{k_B T} + \frac{5}{2} + r\right), \qquad (5)$$

where $r$ is a temperature-independent parameter described by the formula, $r = (d\ln\tau / d\ln E)_{E=E_F}$ with a relaxation time $\tau$.[27] Since the majority carrier in this compound is hole, the $\Delta E$ corresponds to an energy difference between $E_F$ and mobility edge $E_V$ in the hole band. The observed temperature dependence above 250 K can be interpreted on the basis of this equation qualitatively. We compared the observed $S$ with ideal values calculated using Eq. 5 with the $\Delta E$ value obtained by the resistivity data. As a result, we found small quantitative discrepancy between the observed and calculated values. It suggests that the thermoelectric power in the temperature range is affected by not only the thermal activation type band conduction but also somewhat of other mechanisms, *e.g.* effect of minority carriers, polarons, or surviving 2-D VRH regime.

The thermally-induced crossover temperature between the 2-D VRH regime and activation-energy type one is consistent for both the resistivity and the thermoelectric-power measurements. In general, studies on polycrystalline samples sometimes lead to misleading temperature dependence of resistivity of materials owing to external factors such as grain-boundary scattering. However, in this study, we confirmed by the two independent measurements that the crossover of the conduction mechanism occurs at the almost same temperature. In particular, thermoelectric-power data are much reliable because it is not sensitive to the effect of grain-boundary scattering as compared with resistivity data. We therefore concluded that the crossover occurs around 250 K in this compound.

Figure 10 shows temperature dependence of magnetic susceptibility of $(CaOH)_{1.14}CoO_2$. The simple Curie-Weiss-like temperature dependence was observed through the whole temperature range measured. No apparent difference was observed in zero-field cooling data and field cooling ones. Since the present compound is electrically insulating, the magnetic data can be analyzed using the following formula:

$$\chi = \chi_0 + \frac{C}{T - \theta}, \qquad (6)$$

where $\chi_0$ is a temperature-independent component of the magnetic susceptibility, $C$ is the Curie constant, $\theta$ is the Weiss temperature. By fitting the data between 30 K and 300 K with Eq. (6), we obtained the parameters: $\chi_0 = 4.29\times10^{-4}$ emu/mol, $C = 0.0278$ emuK/mol, and $\theta = 1.4$ K. The absolute value of $\theta$ is almost close to zero, suggesting the negligibly weak magnetic interaction between the localized spins. Assuming that the *g*-factor is 2.0, we estimated the effective number of Bohr magnetons to be 0.471 per cobalt ion. This value is generally comparable with the ideal value 0.548 expected on the $Co^{2.9+}$ mix valence state with low-spin configurations, $Co^{3+}$ ($S$=0, $t_{2g}^6 e_g^0$) and $Co^{2+}$ ($S$=1/2, $t_{2g}^6 e_g^1$). It suggests that the Curie term originates in the localized electron spins in the cobalt $e_g$ orbital.

The temperature-independent component, $\chi_0$=4.29×10$^{-4}$ emu/mol, is markedly larger than those of other layered cobalt oxides such as $Tl_\alpha[(Sr_{1-y}Ca_y)O]_{1+x}CoO_2$.[36] To evaluate magnitude of each component included in the $\chi_0$ value, we decomposed the $\chi_0$ into the three terms by the following equation:

$$\chi_0 = \chi_P + \chi_{VV} + \chi_{dia}, \qquad (7)$$

where $\chi_P$ is the Pauli paramagnetic term, $\chi_{VV}$ is the Van-Vleck paramagnetic term derived from the unquenched orbital moments, and $\chi_{dia}$ is the diamagnetic term of closed inner shell of core ions. For the present compound, the $\chi_{dia}$ term can be estimated to be −5.7×10$^{-5}$ emu/mol by summation of diamagnetic values of the constituent ions: −8×10$^{-6}$ emu/mol for $Ca^{2+}$, −1×10$^{-5}$ emu/mol for $Co^{3+}$ and −1.2×10$^{-5}$ emu/mol for $O^{2-}$.[37] The value of the $\chi_{VV}$ term is unknown; however, roughly speaking, it is about 1–2×10$^{-4}$ emu/mol[38], because most of the cobalt oxides have the similar magnitude of the $\chi_{VV}$ value which is almost independent of the kind of materials. Assuming that the $\chi_{VV}$ value of the present compound is 1×10$^{-4}$ emu/mol, we can estimate the $\chi_P$ value to be 3.9×10$^{-4}$ emu/mol. This large Pauli paramagnetic term probably arises from the nature of holes in the cobalt $t_{2g}$ band.

Figure 11 shows temperature dependence of specific heat $C_P$ of $(CaOH)_{1.14}CoO_2$. No apparent anomaly caused by phase transition was observed within the temperature range measured. The $C_P$ at low temperature can be expressed by summation of the electronic specific heat and the lattice specific heat with the following equation:

$$C_P = \gamma T + \frac{12\pi^4}{5}\frac{N_A k_B}{\theta_D^3}T^3, \qquad (8)$$

where $\gamma$ is the Sommerfeld constant, $N_A$ is the Avogadro's number, and $\theta_D$ is the Debye temperature. We obtained the parameters, $\gamma$=3.1 mJ/molK$^2$ and $\theta_D$=488 K, from the extrapolating line shown in Fig. 11. The $\gamma$ obtained is about one order smaller in magnitude than that of other metallic layered cobalt oxides, *e.g.* 31 mJ/molK$^2$ for $Na_{0.7}CoO_2$[8] and 11.0 mJ/molK$^2$ for $Bi_{1.49}Pb_{0.51}Sr_2Co_2O_y$ (Pb-doped $[Bi_{0.87}SrO_2]_2[CoO_2]_{1.82}$).[6] It indicates that the density of states at the Fermi energy level $N(E_F)$ of the present compound is quite smaller than that of the reference compounds. We calculated the $N(E_F)$ value to be 0.66 state/eV per cobalt spin using the formula: $\gamma = (2/3)(\pi k_B)^2 N(E_F)$. Since the majority carrier is hole, the obtained value corresponds to the density of states in the cobalt $t_{2g}$ band. The small $N(E_F)$ value suggests that the Fermi level sits close to the band edge.

The Wilson ratio $R$ ($= (\pi^2/3)(k_B/\mu_B)^2 \times (\chi_P/\gamma)$, $\mu_B$: Bohr magnetons) was estimated to be 2.8. This value is the same level as compared with the Wilson ratio of $Na_{0.7}CoO_2$.[8] The large $R$ suggests presence of strong electron correlation realized in the present compound.

We found that the $C_P/T$ shows the minimum at 10 K and it increases with decreasing temperature below 10 K. Similar upturn of the $C_P/T$ was observed also in the other layered cobalt oxides, $Na_{0.7}CoO_2$ [8], $(Na,Ca)CoO_2$ [39] and $[Bi_{0.87}SrO_2]_2[CoO_2]_{1.82}$ [6], and the reason for the enhancement of the $C_P/T$ has been discussed in terms of spin fluctuation [8,40], Schottky anomaly [39], and premonitory effect of some phase transition realized below 2 K [6] for the respective compounds. For the present compound, at present, the actual reason for the anomalous behavior of the specific heat is not clear and a still open question. Further experimental works are necessary for obtaining the conclusive solution.

As stated above, we obtained the experimental fact that the majority carrier is hole, even though the valence of the cobalt ion is +2.9. In order to explain this fact consistently, we here consider an electronic-band structure realized in $(CaOH)_{1.14}CoO_2$. Figure 12 demonstrates a possible model representing the electronic-band structure near the Fermi energy level $E_F$ for the present compound. The band structure near the $E_F$ principally consists of two energy bands: a cobalt 3d $t_{2g}$ – oxygen 2p hybridized anti-bonding band (= $t_{2g}$-derivative band) and a cobalt 3d $e_g$ – oxygen 2p one (= $e_g$-derivative band). The $t_{2g}$-deribvative band is the main conduction band, in which holes are doped. This situation is similar to those other layer cobalt oxides. [41,42] However, for the present compound, the band edge of the $e_g$-derivative band is lowered below the $E_F$, and electrons are also doped into the $e_g$ band. In other words, both the $e_g$ and $t_{2g}$ bands cross the $E_F$ level, yielding finite density of states at the $E_F$ in the two bands. This band picture is similar to what is expected in typical semimetals, and it can explain the fact that two types of carriers, hole and electron, exist in the cobalt 3d band. The carriers in the tail states (hatched area in the figure) cannot move at low temperature because of the localization due to random potential and insufficiency of doped carriers. The density of states at $E_F$ in the $t_{2g}$-derivative band is small, but it is larger than that in the $e_g$-derivative band.

The electric conduction, thermoelectric power, electronic specific heat, and Pauli paramagnetic term in the magnetic susceptibility are attributed to the holes in the $t_{2g}$ band, while the Curie term in the magnetic susceptibility is attributed to the localized electrons in the $e_g$ band. The hole conduction follows the 2-D VRH regime at low temperature owing to the localization. Near room temperature, some of the holes near $E_F$ are thermally activated to the mobility edge $E_V$ in the $t_{2g}$ band beyond the excitation energy $\Delta E$, and then the dominant conduction mechanism changes from the 2-D VRH regime to the semiconductor-like band conduction one. This causes the crossover of the conduction as observed in both the resistivity and thermoelectric-power measurements. We estimated the $\Delta E$ (=$E_F$−$E_V$) value to be about 0.1 eV from the resistivity data. The total number of electrons in the $t_{2g}$- and $e_g$-derivative bands is 6.1. This comes from the fact that the cobalt valence is +2.9. Based on the proposed electronic-band structure model, we can explain almost all of the experimental results observed.

The electrons in the $e_g$-derivative band are probably less conductive because of its own orbital direction. The cobalt $e_g$ orbital in the $CoO_6$ octahedron spreads out toward the oxygen ions, and is strongly coupled with the oxygen $2p$ orbital. The orbital configurations make edge-shared orthogonal Co–O–Co bonds with a small transfer integral for the electrons. Besides the potential randomness in the lattice system, this effect should make the electrons further less conductive than the holes in the $t_{2g}$ band. This is a possible reason that characteristics of the electrons are hardly detectable by the static transport-property measurements for this compound.

The strong coupling of the $e_g$ orbital with the oxygen ions may cause large electron-phonon interaction such as Jahn-Teller distortion. The electron conduction is strongly affected by the lattice system through the dynamical lattice distortion. The local lattice distortion may trap the electrons into the distorted sites and cause the significant localization of the electrons in the $e_g$ band. This is a kind of state of polarons. The polarons further reduce the mobility for the $e_g$ electrons.

The Jahn-Teller distortion lowers the energy of the $e_g$ band. If large tetragonal distortion of the $CoO_6$ octahedra occurs only at the $Co^{2+}$ ions, it further solves degeneracy of the $e_g$ states and much lowers the electron energy to minimize the total energy of the electron and lattice systems. As a result, the density of states of the $e_g$ band may overlap with that of the $t_{2g}$ band near the $E_F$ as shown in Fig. 12.

We finally estimate the average hopping distance $R_{hop}$ in the 2-D VRH regime. According to Mott *et al.* [27] and Brenig *et al.* [35], in the case of 2-D VRH regime, the $R_{hop}$ and $T_0$ in Eq. (1) can be written by

$$R_{hop} = \left( \frac{C'}{\alpha N'(E_F) k_B T} \right)^{\frac{1}{3}} \quad (9)$$

and

$$T_0 = \frac{C'' \alpha^2}{N'(E_F) k_B}, \quad (10)$$

where $N'(E_F)$ is the density of states in unit area per cobalt ion, $\alpha$ is the decay constant of the localized wave function. In these equations, we set the coefficients as follows: $C'=1/\pi$ and $C''=3^3/\pi$.

We obtained the $T_0$ value to be $4.7\times10^5$ K by fitting the resistivity data in the temperature range between 130 K and 270 K with Eq. (1), and the $N(E_F)$ value to be $1.9\times10^{15}$ state/eVcm$^2$ using the $N(E_F)$ value obtained from the specific-heat measurement. Substituting the calculated $T_0$ and $N'(E_F)$ values into Eq. (10), we obtained the $\alpha$ value to be 0.95 Å$^{-1}$. Subsequently, substituting the $\alpha$ and $N'(E_F)$ values into Eq. (9), we obtained the $R_{hop}$ values as a function of $T$. The typical $R_{hop}$ values calculated are 5.4 Å at 130 K and 4.2 Å at 270 K. These values are longer than the nearest-neighbor distance, ≈2.8 Å, between adjacent cobalt ions. This indicates that the nearest-neighbor hopping is forbidden within the temperature range between 130 K and 270 K. This is a necessary condition for realizing the VRH regime in the present compound. We are therefore confident that the VRH regime is a suitable model for representing the conduction within the temperature range.

**IV. SUMMARY**

A new cobalt oxide (CaOH)$_{1.14}$CoO$_2$ has been synthesized under high pressure. This compound is a kind of composite crystal consisting of two interpenetrating subsystems: CdI$_2$-type CoO$_2$ layers and rock-salt-type Ca$_2$(OH)$_2$ blocks. The two subsystems have aperiodicity along the *a*-axis, and are alternately stacked along the *c*-axis. The structure is modulated owing to interaction between the subsystems. Hydrogen atoms are situated in the rock-salt-type block as OH$^-$ ions. This compound is the first cobalt oxide which has double rock-salt-type CaOH atomic layers in the structure.

The resistivity and thermoelectric power measurements revealed that the conduction mechanism follows two-dimensional variable-range-hopping regime below 220 K, and that it undergoes crossover to activation-energy type band conduction near room temperature. The positive sign of the Seebeck coefficient clearly indicated that the majority carrier type is hole through the temperature range observed. The crossover of the conduction mechanism seems to be related to thermal excitation of localized holes near Fermi energy level to the mobility edge in the hole band. For the magnetic susceptibility and specific-heat measurements, we found the large Pauli paramagnetic component of ≈$3.9\times10^{-4}$ emu/mol and the comparatively small Sommerfeld constant of ≈3.1 mJ/molK$^2$, resulting in the large Wilson ratio of ≈2.8. This indicates presence of strong electron correlation in this compound.

In order to explain the experimental results: (i) majority carrier is hole, and (ii) the valence of the cobalt ion is +2.9, we suggested a semimetal-like electronic-band structure model. The band structure near Fermi energy level consists of the cobalt $t_{2g}$ band overlapping with the cobalt $e_g$ band. Both the $t_{2g}$ and $e_g$ band cross the Fermi energy level near the band edge, yielding finite density of states in the two bands. The electric conduction, thermoelectric power, Pauli paramagnetic term in the magnetic susceptibility and electronic specific heat are mainly attributed to the holes in the $t_{2g}$ band, while the Curie-Weiss behavior in the magnetic susceptibility is due to the electrons in the $e_g$ band.

ACKNOWLEDGEMENTS

We thank Drs. Akaishi, Taniguchi and Kanke of NIMS for a lot of helpful advice on the high-pressure experiments. One of the authors (MS) thanks Dr. Sakurai of NIMS for fruitful discussion on synthesis and physical properties. This research work was partly supported by the Superconducting Materials Research Project administrated by the Ministry of Education, Culture, Sports, Science and Technology of Japan.

--------------------------------------------------------------------------------------------------

* To whom correspondence should be addressed. E-mail: ISOBE.Masaaki@nims.go.jp. Fax. +81-29-860-4674.

conceive that this is not intrinsic property of the present phase and is due to the problems of the measurement system on measuring the high-resistance sample. In this paper, we do not discuss the resistivity data in the temperature range.

**FIGURES**

FIG. 1    Schematic representation of crystal structure of layered cobalt oxides. The $n$ indicates the number of atomic planes in the blocking layer situated between the $CoO_2$ layers. The $n=2$ compound (($CaOH)_{1.14}CoO_2$) is a new member of the series reported in the present work.

FIG. 2    Powder x-ray diffraction patterns for the washed-sample of $(CaOH)_{1.14}CoO_2$; the inset is a whole $2\theta$ range pattern, while the main panel is an enlargement of the pattern. The arrows in the main panel indicate diffraction-peak positions of $Ca(OH)_2$.

FIGS. 3    Electron diffraction (ED) patterns projected along the [001] (a), [010] (b), and [100] (c) directions for $(CaOH)_{1.14}CoO_2$.

FIG. 4    High-resolution transmission electron microscopy (HRTEM) image projected along the direction perpendicular to the $c$-axis for $(CaOH)_{1.14}CoO_2$.

FIG. 5    Possible crystal structure model of $(CaOH)_{1.14}CoO_2$. The rectangles indicate unit lattices for the two subsystems. The open circles and squares represent the cobalt atoms situated at different positions along the projected coordinate. This drawing is idealized as if its structure was not modulated at all. (A part of this figure was drawn with a computer software VENUS developed by R. A. Dilanian and F. Izumi).

FIG. 6    X-ray diffraction pattern simulated on the basis of the proposed crystal structure model for $(CaO)_{1.14}CoO_2$. The tick marks indicate positions arrowed Bragg reflections. For comparison, the observed x-ray diffraction pattern is also displayed at the upper panel in the figure.

FIG. 7    Raman scattering spectrum for a crystalline grain in the washed-sample for $(CaOH)_{1.14}CoO_2$.

FIGS. 8    Temperature dependence of electrical resistivity of $(CaOH)_{1.14}CoO_2$: (a) linear-logarithm plot and (b) logarithmic-derivative plot. In Fig. (b), the slope of the fit line gives $\nu$ in Eq. (1).

FIG. 9    Seebeck coefficient $S$ of $(CaOH)_{1.14}CoO_2$ as a function of temperature $T$. The inset shows a part of enlargement of a logarithm-logarithm plot of the data. The solid line indicates $T^{0.34}$ dependence of $S$.

FIG. 10　Temperature dependence of magnetic susceptibility $\chi$ per cobalt ion for $(CaOH)_{1.14}CoO_2$. The scale of the right vertical axis is for inverse susceptibility $1/(\chi-\chi_0)$, where $\chi_0$ (=$4.29\times10^{-4}$ emu/mol) is a temperature-independent component in the magnetic susceptibility.

FIG. 11　Temperature dependence of specific heat $C_P$ per cobalt ion for $(CaOH)_{1.14}CoO_2$; the inset indicates a $C_P$–$T$ linear-scale plot, while the main panel indicates a $C_P/T$–$T^2$ plot in the low temperature range. The linear line indicates a data-fit line by Eq. (8) with the parameters: $\gamma$=3.1 mJ/mol K$^2$ and $\theta_D$=488 K.

FIG. 12　Electronic-band structure model near Fermi energy level $E_F$ for $(CaOH)_{1.14}CoO_2$. Both the $e_g$ and $t_{2g}$ bands cross the $E_F$, yielding finite density of states at $E_F$ in the two bands. The $E_V$ indicates mobility edge of the $t_{2g}$ band. Hatched area indicates tail states in which carriers are localized by random potential. The electrons in the $e_g$ band are further less conductive than the $t_{2g}$ holes owing to orthogonality of the direction of the cobalt $e_g$ orbitals. (See text.)

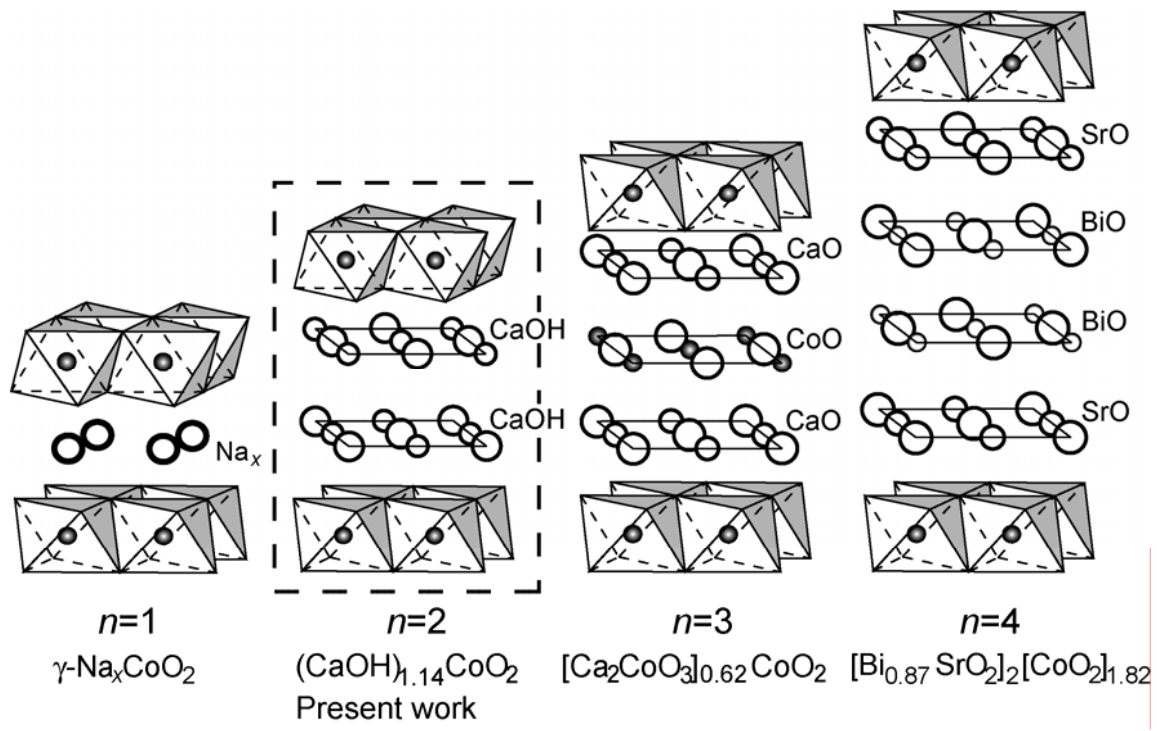

FIG. 1 M. Shizuya, *et al.*

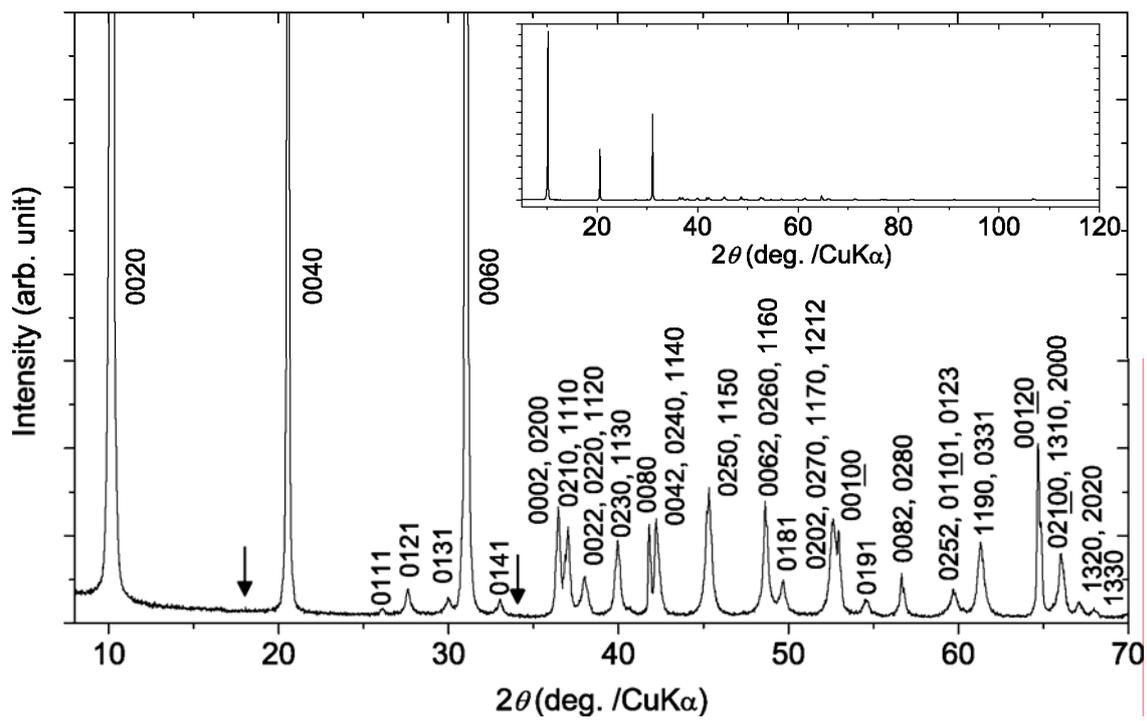

FIG. 2  M. Shizuya, *et. al.*

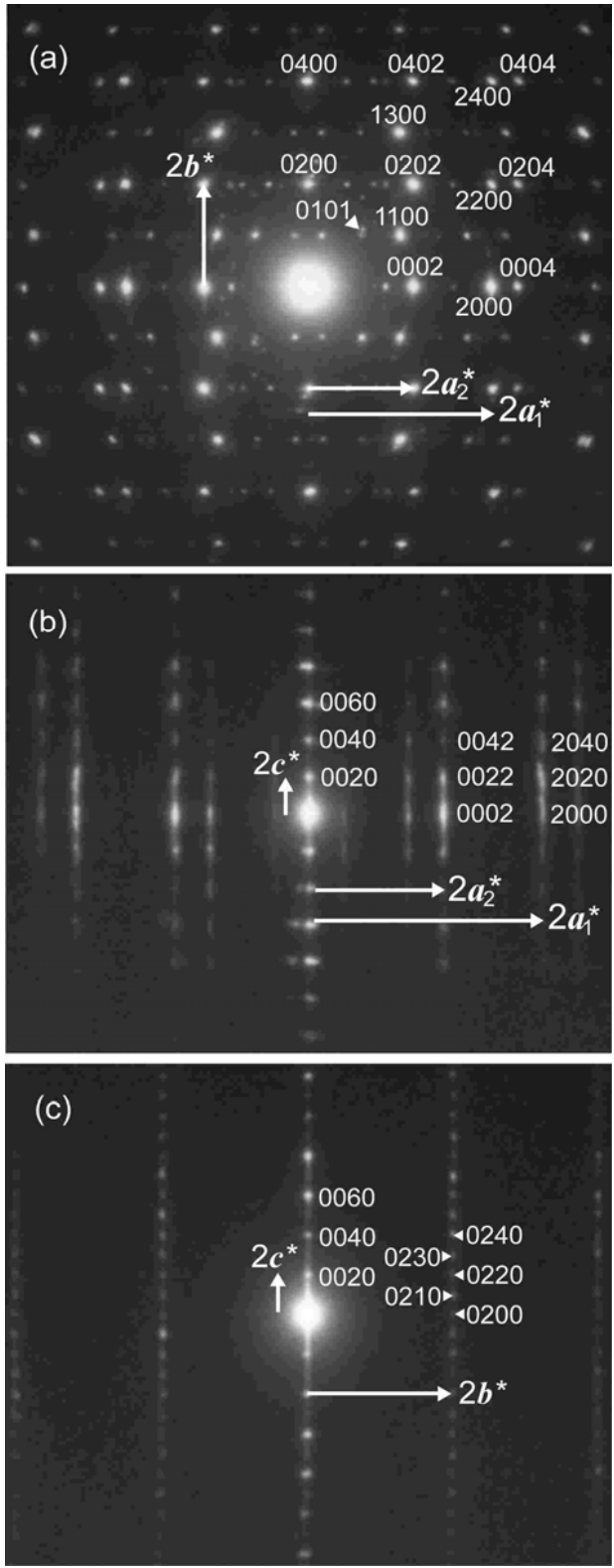

FIGS. 3   M. Shizuya, *et al.*

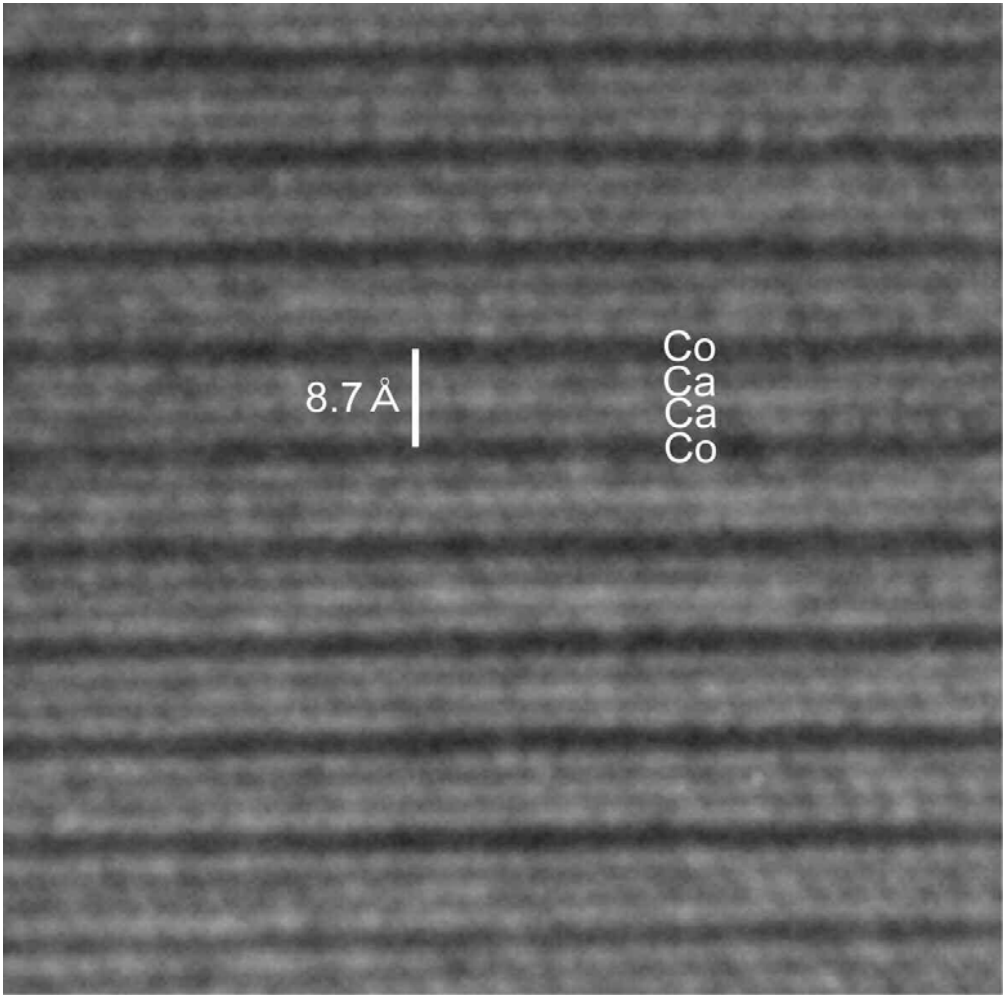

FIG. 4   M. Shizuya, *et al*.

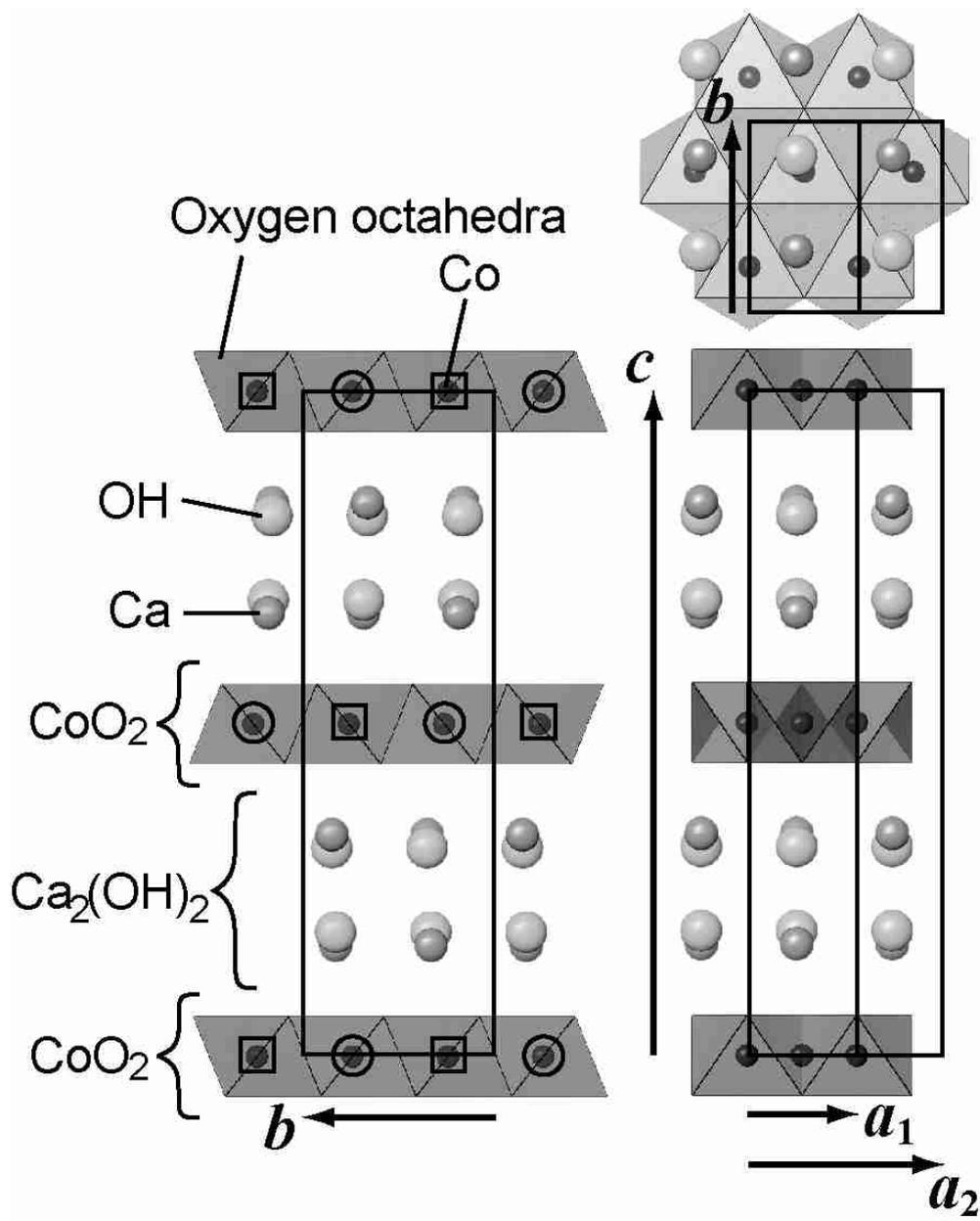

FIG. 5  M. Shizuya, *et al.*

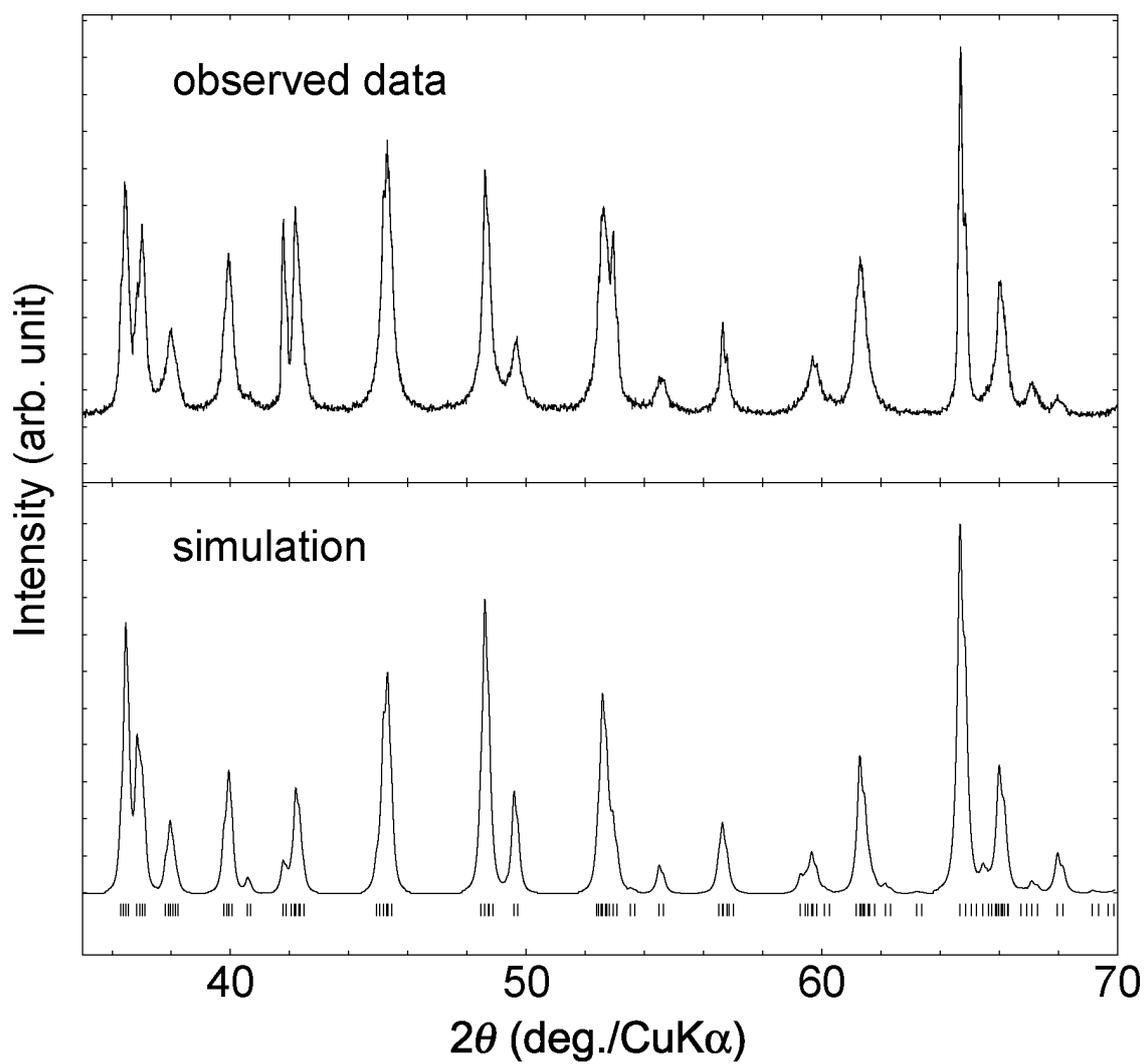

FIG. 6  M. Shizuya, *et al.*

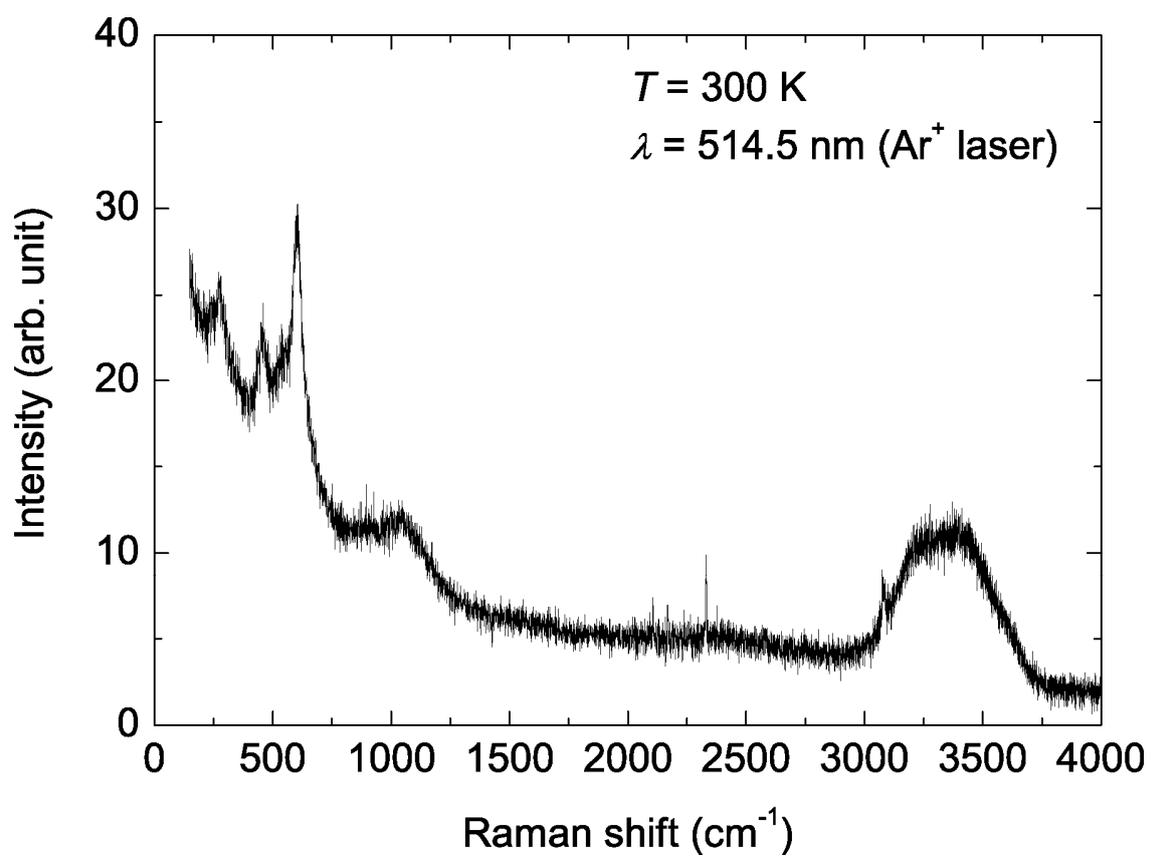

FIG. 7  M. Shizuya, *et al.*

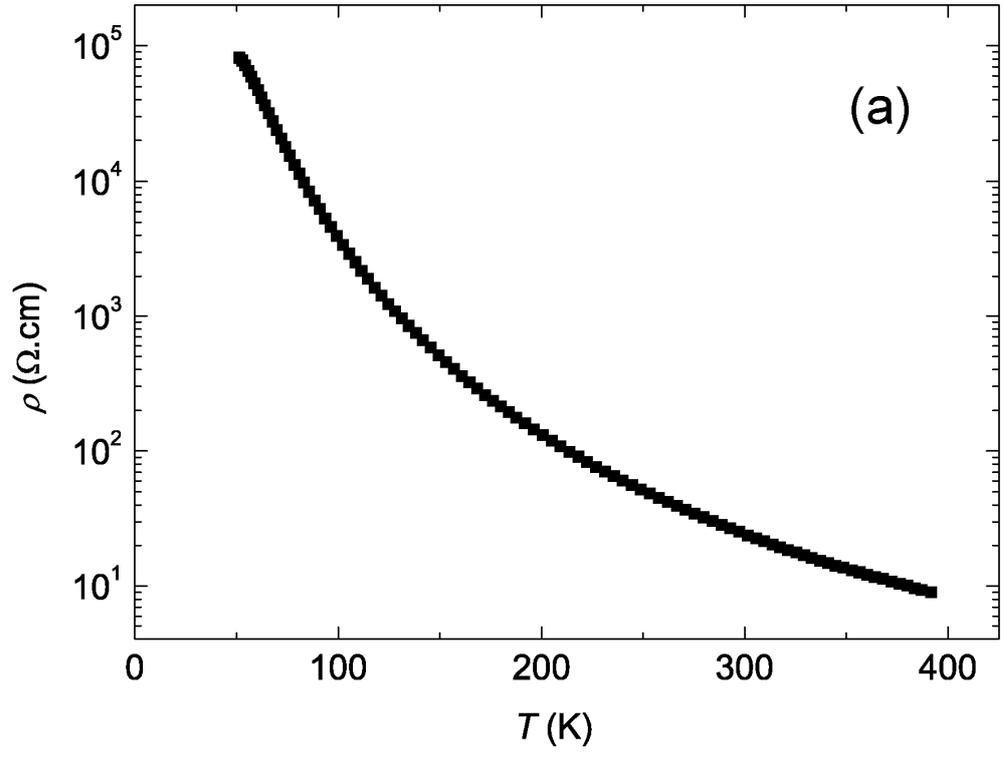

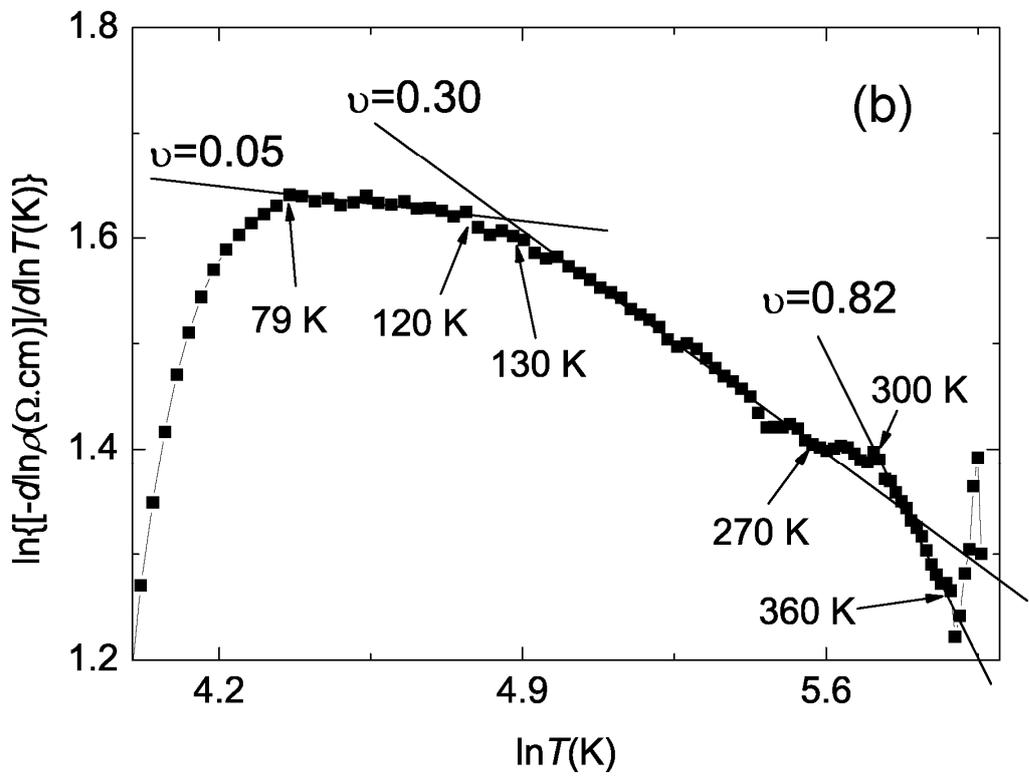

FIGS. 8  M. Shizuya, *et al.*

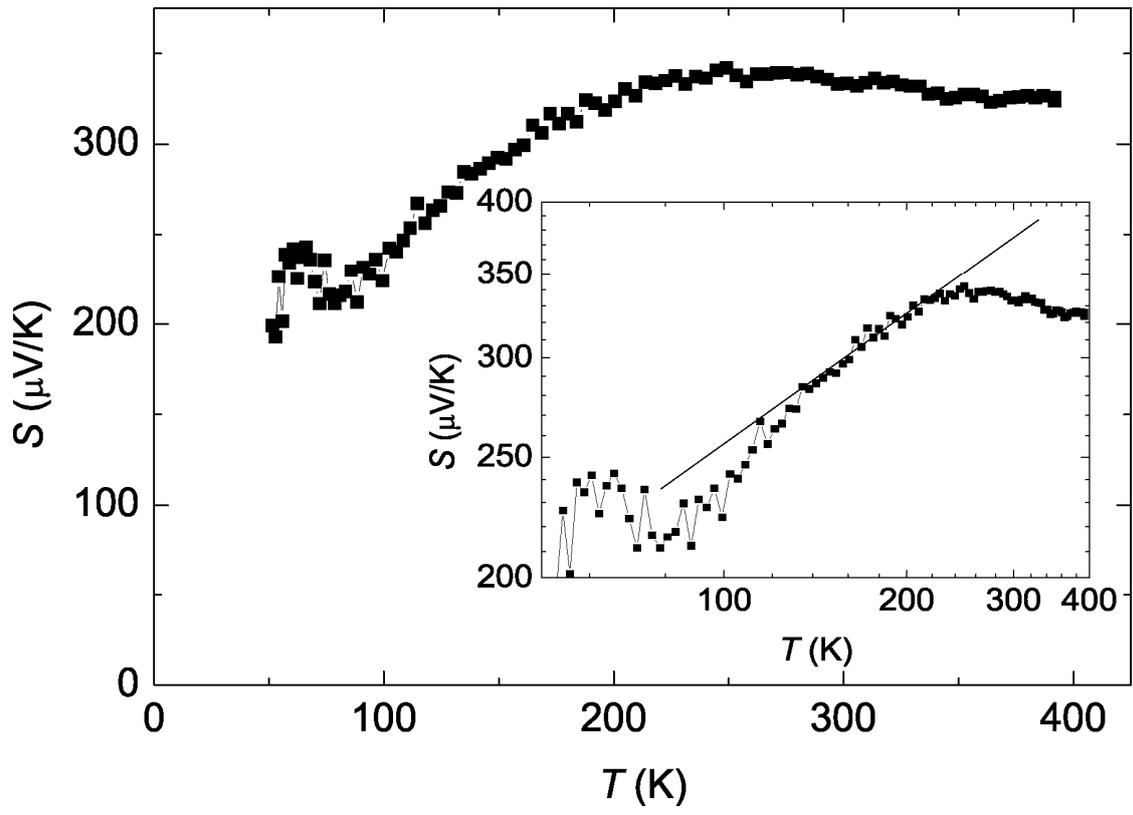

FIG. 9    M. Shizuya, *et al.*

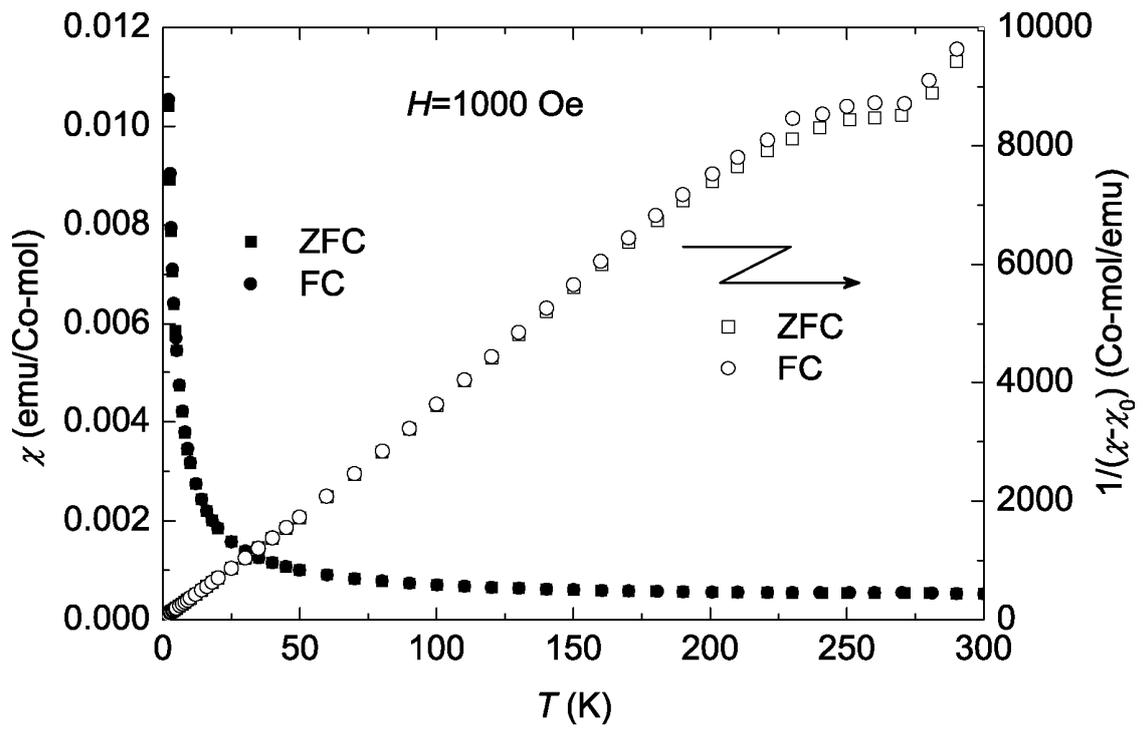

FIG. 10  M. Shizuya, *et al.*

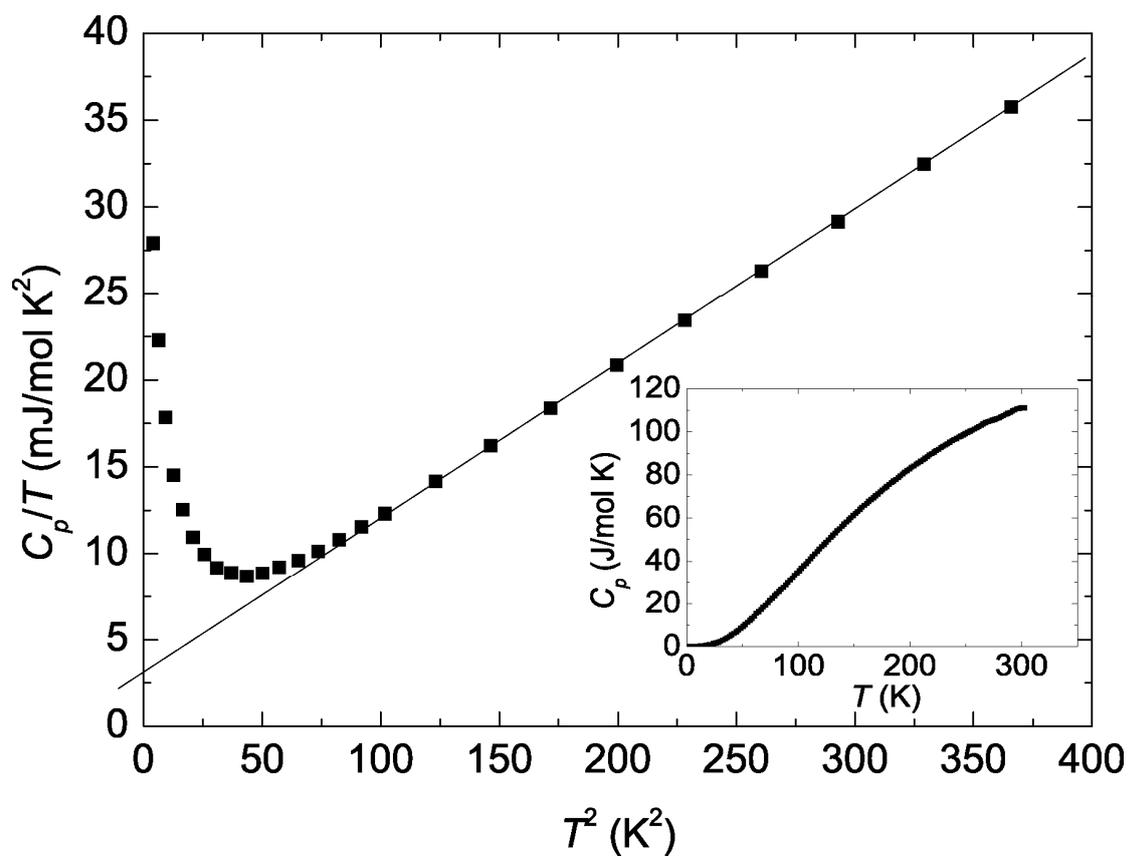

FIG. 11  M. Shizuya, *et al.*

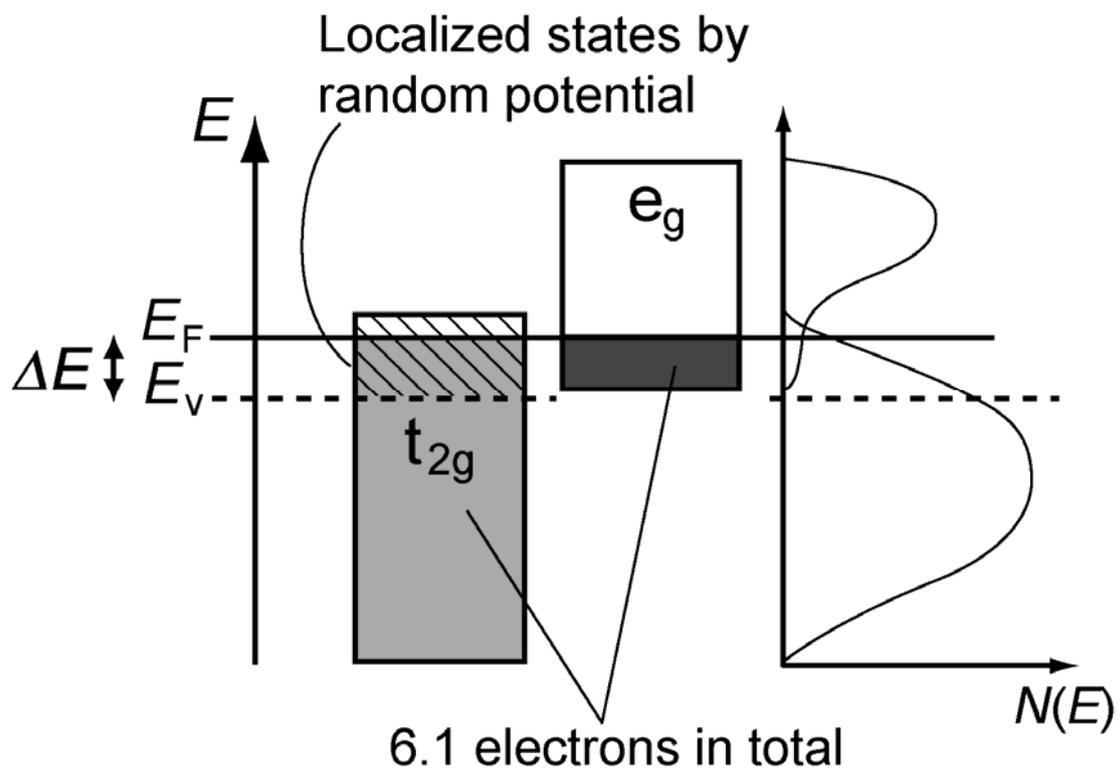

FIG. 12   M. Shizuya, et al.